\documentclass[preprint,preprintnumbers,showpacs,aps,amssymb]{revtex4}

\usepackage{graphicx}
\usepackage{bm}
\usepackage{amsmath}

\def\calL{{\cal L}}
\def\calO{{\cal O}}
\def\calU{{\cal U}}

\def\dU{d_{\calU}}

\def\zbar{{\bar z}}

\def\hn{h^{(n)}}
\def\tn{t^{(n)}}

\def\AdS4N{{\rm AdS}_{4+N}}
\def\MP{M_{\rm Pl}}

\def\nn{\nonumber}

\begin{document}
\title{Ungravity realization in fractional extra dimensions}
\author{Jong-Phil Lee}
\email{jplee@kias.re.kr}
\affiliation{Korea Institute for Advanced Study, Seoul 130-722, Korea}
\preprint{KIAS-P09055}

\begin{abstract}
Ungravity by tensor unparticles is realized in $\AdS4N$ space through 
deconstruction.
It is shown that ungravity is equivalent to the gravity in extra dimensions.
There is a close relation between the scaling dimension of the unparticle
operator and the number of extra dimensions.
Consequently it is possible to discover the fractional extra dimentions,
which would be a stringent signal for unparticles.
\end{abstract}

\maketitle
\section{Introduction}
With the reoperation of the Large Hadron Collider (LHC) at CERN very recently,
we are now entering a new era of physics in history.
The LHC will unveil many mysteries of high energy physics such as electroweak
symmetry breaking, dark matter, and new symmetries, to name a few.
It seems quite true that the standard model (SM) of particle physics is only
an effective theory at low energy, and there must be some new physics behind it.
Many kinds of new physics --- supersymmetry or extra dimensions, etc. --- 
involve some new sets of {\em particles}.
But recently a totally different type of new physics was suggested by Georgi 
\cite{Georgi}.
In this scenario, there is a scale-invariant hidden sector which couples to the
SM particles very weakly.
When seen at low energy, the hidden sector behaves in different ways from 
ordinary particles, hence dubbed as "unparticles."
\par
Consider a ultraviolet (UV) theory in the hidden sector with the 
infrared (IR)-stable fixed point.
The theory interacts with the SM sector at a scale of $M_\calU$.
Below $M_\calU$, the interaction between a UV operator $\calO_{\rm UV}$ and
an SM operator $\calO_{\rm SM}$ is described as 
$\calO_{\rm SM}\calO_{\rm UV}/M_\calU^{d_{\rm SM}+d_{\rm UV}-4}$.
Here $d_{\rm UV(SM)}$ is the scaling dimension of $\calO_{\rm UV(SM)}$.
When the scale goes down via the renormalization flow, a scale 
$\Lambda_\calU$ appears through the dimensional transmutation where the
scale invariance emerges.
Below $\Lambda_\calU$ the theory is matched onto the above interaction with
the new unparticle operator $\calO_\calU$ as
\begin{equation}
C_\calU\frac{\Lambda_\calU^{d_{\rm UV}-\dU}}{M_\calU^{d_{SM}+d_{\rm UV}-4}}
\calO_{SM}\calO_{\calU}~,
\end{equation}
where $\dU$ is the scaling dimension of $\calO_\calU$ and $C_\calU$ is the
matching coefficient.
Because of the scale invariance, $\dU$ can have nontrivial values.
This unusual behavior is reflected on the phase space of $\calO_\calU$.
To see it, consider the spectral function of the unparticle which is given by
the two-point function of $\calO_\calU$:
\begin{eqnarray}
\rho_\calU(P^2)&=&\int d^4x~e^{iP\cdot x}
\langle 0|\calO_\calU(x)\calO_\calU^\dagger(0)|0\rangle\nn\\
&=&
A_{\dU}\theta(P^0)\theta(P^2)(P^2)^{\dU-2}~,
\label{rhoU}
\end{eqnarray}
where
\begin{equation}
A_{\dU}=\frac{16\pi^2\sqrt{\pi}}{(2\pi)^{2\dU}}
\frac{\Gamma(\dU+\frac{1}{2})}{\Gamma(\dU-1)\Gamma(2\dU)}~,
\end{equation}
is the normalization factor.
The corresponding phase space is
\begin{equation}
d\Phi_\calU(P)=\rho_\calU(P^2)\frac{d^4P}{(2\pi)^4}
=A_{\dU}\theta(P^0)\theta(P^2)(P^2)^{\dU-2}\frac{d^4P}{(2\pi)^4}~.
\end{equation}
Since $\dU$ can be any real number, it looks like a phase space for a fractional
number of particles. 
\par
Till now there have been a lot of investigations about unparticles in every
respect \cite{Kingman, U}.
Among them is the so called ungravity \cite{Goldberg,Das,Mureika}.
Ungravity is induced by a traceless tensor unparticle $\calO_{\mu\nu}$ 
with the interaction
\begin{equation}
\kappa_*\frac{1}{\Lambda_\calU}\sqrt{g}T^{\mu\nu}\calO_{\mu\nu}~,
\end{equation}
where $\kappa_*=\Lambda_\calU^{-1}(\Lambda_\calU/M_\calU)^{d_{\rm UV}}$.
The most important result of ungravity is the power law correction to the
Newtonian gravitational potential, of type $\sim (1/r)^{2\dU-1}$.
This type of power law correction reminds one of the extra dimensional 
scenarios \cite{ADD,RS}.
Typically for extra $N$ dimensions, the Newtonian gravity potential gets
corrections $\sim (1/r)^{N+1}$ \cite{ADD,Nima,Garriga,Csaki}.
In fact, there are much stronger motivations to relate unparticles to extra
dimensions.
As already pointed out in \cite{Kingman}, the unparticle and the Kaluza-Klein
(KK) states of extra dimensions share analogous phase space integrals.
The integral over mass spectrum of KK states behaves as $(m^2)^{N/2-1}dm^2$.
Comparing with Eq.\ (\ref{rhoU}), one has $\dU=1+N/2$, which is consistent with
the results from comparing the gravitational potential corrections.
\par
Furthermore, there is a transparent way of realizing unparticles, known as 
deconstruction \cite{Stephanov}, which looks much like the KK decomposition.
In this scheme the unparticle is described by an infinite tower of particles
with vanishing masses.
A continuous spectrum of unparticles is simulated by a descrete sum over
deconstructing states, which comes to an integral in the vanishing mass limit.
One way of explicit realization of deconstruction is to use AdS/CFT
correspondence \cite{Maldacena} to build a 5-dimensional field theory.
But it is also possible to build flat $4+N$ dimensional theory for 
deconstructing unparticles \cite{jplee}.
In the framework of \cite{jplee}, it can be easily shown that the
spectral function shifts as $\rho_\calU(P^2)\to\rho_\calU(P^2-\mu^2)$ 
when the scale invariance is broken by a new scale $\mu^2$.
\par
In this paper, we try to construct $4+N$ dimensional theoy in anti-de Sitter
(AdS) space to realize tensor unparticles by decosntruction.
The main idea is that the ungravity effects by tensor unparticles are 
{\em equivalent} to the effects of excited KK modes in $\AdS4N$.
Again one gets a relation between the scaling dimension of the unparticle
operator and the number of extra dimensions, $\dU=1+N/2$.
Since $\dU$ does not have to be integers in principle, it is possible that
we would confront with the fractional extra dimensions (FXD)!
In other words, if we find signals at the LHC telling that the number of 
extra dimensions is not an integer, it would be a strong evidence for
unparticles.
\par
In the next section, it is given how to deconstruct ungravity.
In sec. III, gravity in $\AdS4N$ is introduced and related to the 
deconstructed ungravity of the previous section.
Section IV contains discussions and conclusions.
\section{Deconstructing ungravity}
Consider first the ungravity due to the tensor unparticle operator $\calO_{\mu\nu}$:
\begin{equation}
\calL_{\rm ung}\equiv
\frac{\kappa_*}{\Lambda_\calU^{\dU-1}}\sqrt{g}T^{\mu\nu}\calO_{\mu\nu}~.
\end{equation}
The two-point function of $\calO_{\mu\nu}$ is
\begin{equation}
\langle 0|\calO_{\mu\nu}(x)\calO_{\alpha\beta}^\dagger(0)|0\rangle
=\int\frac{d^4P}{(2\pi)^4}~e^{-iP\cdot x}~\rho_{\mu\nu\alpha\beta}(P^2)~,
\end{equation}
where the spectral function $\rho_{\mu\nu\alpha\beta}(P^2)$ is given by
\begin{equation}
\rho_{\mu\nu\alpha\beta}(P^2)=A_{\dU}\theta(P^0)\theta(P^2)(P^2)^{\dU-2}
\Pi_{\mu\nu\alpha\beta}(P)~.
\end{equation}
The tensor structure of $\rho_{\mu\nu\alpha\beta}$ is encoded in
$\Pi_{\mu\nu\alpha\beta}$.
On the other hand, the structure of the two-point function is fixed by the
scale invariance.
In general, one can put \cite{GIR}
\begin{eqnarray}
\langle 0|\calO_{\mu\nu}(x)\calO_{\alpha\beta}^\dagger(0)|0\rangle
&=&
c_T\frac{1}{(2\pi)^2}\frac{1}{(x^2)^{\dU}}\left\{\left[
   I_{\mu\alpha}(x)I_{\nu\beta}(x)+\mu\leftrightarrow\nu\right]
   -\frac{1}{2}g_{\mu\nu}g_{\alpha\beta}\right\}\nn\\
&=&
c_T\frac{\Gamma(2-\dU)}{4^{\dU-1}\Gamma(\dU+2)}
\int\frac{d^4P}{(2\pi)^4}~e^{-iP\cdot x}~(P^2)^{\dU-2}T_{\mu\nu\alpha\beta}(P)~,
\end{eqnarray}
where
\begin{equation}
I_{\mu\nu}(x)\equiv g_{\mu\nu}-2\frac{x_\mu x_\nu}{x^2}~.
\end{equation}
Here the tensor structure is encoded in $T_{\mu\nu\alpha\beta}$ as
\begin{eqnarray}
T_{\mu\nu\alpha\beta}(P)&=&
\dU(\dU-1)(g_{\mu\alpha}g_{\nu\beta}+\mu\leftrightarrow\nu)
+\left[2-\frac{\dU}{2}(\dU+1)\right]g_{\mu\nu}g_{\alpha\beta}\nn\\
&&
-2(\dU-1)(\dU-2)\left(g_{\mu\alpha}\frac{P_\nu P_\beta}{P^2}
   +g_{\mu\beta}\frac{P_\nu P_\alpha}{P^2}+\mu\leftrightarrow\nu\right)\nn\\
&&
+4(\dU-2)\left(g_{\mu\nu}\frac{P_\alpha P_\beta}{P^2}
   +g_{\alpha\beta}\frac{P_\mu P_\nu}{P^2}\right)
+8(\dU-2)(\dU-3)\frac{P_\mu P_\nu P_\alpha P_\beta}{(P^2)^2}~.
\end{eqnarray}
Combining the two expressions one can fix
\begin{eqnarray}
\Pi_{\mu\nu\alpha\beta}&=&T_{\mu\nu\alpha\beta}~,\\
c_T&=&\frac{4^{\dU-1}\Gamma(\dU+2)}{\Gamma(2-\dU)}~A_{\dU}
=
-\frac{8\pi^3}{(2\pi)^{2\dU}}\dU(\dU+1)\frac{\sin\pi\dU}{\pi}~.
\end{eqnarray}
The propagator of the tensor unparticle is
\begin{eqnarray}
\Delta_{\mu\nu\alpha\beta}(P)&=&
\int d^4x~e^{iP\cdot x}~
\langle 0|T~\calO_{\mu\nu}(x)\calO_{\alpha\beta}^\dagger(0)|0\rangle\nn\\
&=&
\frac{1}{2\pi}\int dM^2\frac{i}{P^2-M^2+i\epsilon}
\rho_{\mu\nu\alpha\beta}(M^2)~\nn\\
&=&
\frac{A_{\dU}}{2\sin(\pi\dU)}(-P^2)^{\dU-2}T_{\mu\nu\alpha\beta}~.
\end{eqnarray}
Note that the propagator above is different from that of \cite{Goldberg,Das}
in tensor structure.
Consequently, the resulting ungravity effect on the modification of Newtonian
gravity must be changed.
\par
Now we deconstruct the unparticle operator $O_{\mu\nu}$ into the infinite
tower of states $|\lambda_n\rangle$ with infinitesimal mass.
One can write
\begin{equation}
O_{\mu\nu}\equiv\sum_n F_n \tn_{\mu\nu}~,
\label{decon}
\end{equation}
where
\begin{equation}
\epsilon_{\mu\nu}=\langle 0|\tn_{\mu\nu}(0)|\lambda_n\rangle
\end{equation}
is the polarization tensor.
\par
With the deconstruction, the spectral function is given by
\begin{equation}
\rho_{\mu\nu\alpha\beta}=2\pi\sum_\lambda\delta(P^2-p_\lambda^2)F_\lambda^2
~P_{\mu\nu\alpha\beta}(p_\lambda)~,
\end{equation}
where
\begin{equation}
P_{\mu\nu\sigma\rho}=\frac{1}{2}(P_{\mu\sigma}P_{\nu\rho}+
P_{\mu\rho}P_{\nu\sigma}-\alpha P_{\mu\nu}P_{\sigma\rho})~,
\end{equation}
and $P_{\mu\nu}(p)\equiv -\eta_{\mu\nu}+p_\mu p_\nu/p^2$.
For a massive graviton, one has $\alpha=2/3$.
The corresponding propagator is
\begin{equation}
\Delta_{\mu\nu\alpha\beta}(P)=
\sum_\lambda\frac{iF_\lambda^2}{P^2-p_\lambda^2+i\epsilon}
~P_{\mu\nu\alpha\beta}(p_\lambda)~.
\end{equation}
The "decay constant" $F_\lambda$ is matched as
\begin{equation}
2\pi\sum_\lambda\delta(P^2-p_\lambda^2)F_\lambda^2~
P_{\mu\nu\alpha\beta}(p_\lambda)
=
A_{\dU}\theta(P^0)\theta(P^2)(P^2)^{\dU-2}
\Pi_{\mu\nu\alpha\beta}(P)~.
\label{matchF}
\end{equation}
The ungravity Lagrangian $\calL_{\rm ung}$ is now
\begin{equation}
\calL_{\rm ung}=
\frac{\kappa_*}{\Lambda_\calU^{\dU-1}}\sqrt{g}T^{\mu\nu}
\sum_n F_n\tn_{\mu\nu}~.
\label{Lung}
\end{equation}
\par
The presence of $\calL_{\rm ung}$ modifies the Newtonian gravitational 
potential through the exchange of unparticles.
The amount of modification is \cite{Goldberg}
\begin{equation}
V_\calU(r)=-\frac{m_1 m_2 G}{r}~\left(\frac{R_G}{r}\right)^{2\dU-2}~,
\label{VU}
\end{equation}
where
\begin{equation}
R_G=\frac{1}{\pi\Lambda_\calU}(\kappa_* M_{\rm Pl})^{1/(\dU-1)}
\left[\frac{2(2-\alpha)}{\pi}
\frac{\Gamma(\dU+1/2)\Gamma(\dU-1/2)}{\Gamma(2\dU)}\right]^{1/(2\dU-2)}~.
\end{equation}
\section{Gravity in $\AdS4N$}
The form of Eq.\ (\ref{decon}) or (\ref{Lung}) reminds one of the KK
decomposition of higher dimensional gravitions.
As a concrete example, we consider $4+N$-dimensional AdS space with the metric
(in the Poincare parametrization) \cite{Nima}
\begin{equation}
ds^2_{4+N}=\frac{L^2}{z^2}(\eta_{\mu\nu}dx^\mu dx^\nu+d{\vec w}^2_{N-1}
+dz^2)~,
\end{equation}
After some reparametrizations one can arrive at
\begin{equation}
ds^2_{4+N}=\Omega^2\left(\eta_{\mu\nu} dx^\mu dx^\nu
+\sum_{i=1}^N(d\zbar^i)^2\right)~,
\end{equation}
where
\begin{eqnarray}
\Omega&\equiv&\frac{1}{k\sum_j|\zbar^j|+1}~,\nn\\
k&\equiv&\frac{1}{\sqrt{N}L}~.
\end{eqnarray}
Here the new coordinates $\zbar^j$ are obtained by a rotation such that
$z=\sum_{j=1}^N\zbar^j/\sqrt{N}$.
The linearized perturbation $h_{\mu\nu}(x,\zbar)$ around $\eta_{\mu\nu}$
satisfies the field equation \cite{Nima}
\begin{equation}
\left[\frac{1}{2}\Box_4-\frac{1}{2}\nabla^2_\zbar+V(\zbar)\right]
{\hat h}=0~,
\end{equation}
where ${\hat h}=\Omega^{(N+2)/2} h$ with $\mu\nu$ indices dropped, and
\begin{equation}
V(\zbar)=\frac{N(N+2)(N+4)k^2}{8}\Omega^2
-\frac{(N+2)k}{2}\Omega~\sum_j\delta(\zbar^j)~.
\end{equation}
If we decompose ${\hat h}(x,\zbar)=e^{ipx}{\hat\psi}(\zbar)$,
the 4D mass $m_\lambda=\sqrt{p^2}$ for the $\lambda$-th level is
 determined through
\begin{equation}
\left(-\frac{1}{2}\nabla_\zbar^2+V(\zbar)\right){\hat\psi}_\lambda
=\frac{1}{2}m_\lambda^2{\hat\psi}_\lambda~.
\label{DE}
\end{equation}
For a test mass $M$ located at $x=\zbar=0$,
the gravitational potential $U(r=|{\vec x}|)$ is
\begin{eqnarray}
\frac{U(r)}{M}&=&\sum_\lambda G_{N(4+N)}~|{\hat\psi}_\lambda(0)|^2~
   \frac{e^{-m_\lambda r}}{r}\nn\\
&\sim&
G_{N(4)}\frac{1}{r}
+\sum_{\rm continuum}G_{N(4+N)}~|{\hat\psi}_\lambda(0)|^2~
  \frac{e^{-m_\lambda r}}{r}~,
\end{eqnarray}
where $G_{N(4)}\sim G_{N(4+N)}/L^N$.
The continuum contribution produces the $(4+N)$-dimensional potential
$\sim G_{N(4+N)}/r^{1+N}$.
\par
At this stage, one can find a strong similarity between ungravity and the
$(4+N)$-dimensional graivty.
The action for the $(4+N)$-dimensional gravity is
\begin{equation}
S\sim\int d^{4+N}x\sqrt{g_{4+N}}M_*^{N+2}R_{4+N}
\sim\int d^4x\sqrt{g_4}T^{\mu\nu}\int d^N\zbar\sqrt{g_N} M_*^{N}h_{\mu\nu}~,
\label{S}
\end{equation}
which is the same in form as Eq.\ (\ref{Lung}) because
$h(x,\zbar)\sim\sum_\lambda e^{ip_\lambda x}{\hat\psi}_\lambda(\zbar)$.
The resulting modification of the Newtonian potential is $\sim 1/r^{1+N}$ in
$4+N$ dimensional theory while $\sim 1/r^{2\dU-1}$ for ungravity, as given
in Eq.\ (\ref{VU}).
Thus one can identify $1+N=2\dU-1$, or
\begin{equation}
N=2(\dU-1)~.
\end{equation}
Naively, we expect (neglecting the overall dimension for the moment)
\begin{equation}
\calO_{\mu\nu}=\sum_n \tn_{\mu\nu}F_n
\sim \sum_{n\neq 0} \hn_{\mu\nu}(x)
\left[M_*^N\int d^N\zbar\sqrt{g_N}\psi_n(\zbar)\right]~.
\end{equation}
The zero mode of KK excitation is excluded since it is just
the usual 4-dimensional gravity.
The continuum modes of KK states correspond to the deconstructed continuous
states of the unparticle.
Since $M_*\sim L^{-1}$, the factor of $M_*^N$ in front of the $\zbar$-integral
plays the role of normalization with the "volume" $L^N$.
More specifically, if we consider the spectral function of $\calO_{\mu\nu}$,
\begin{eqnarray}
\rho_{\mu\nu\alpha\beta}(P)&=&
\int d^4x e^{iP\cdot x}
\langle 0|\calO_{\mu\nu}(x)\calO_{\alpha\beta}^\dagger(0)|0\rangle\nn\\
&\sim&
\int d^4x e^{iP\cdot x}\sum_{n,m}
\langle 0|\hn_{\mu\nu}(x)\chi_n
h^{(m)\dagger}_{\alpha\beta}(0)\chi_m^\dagger|0\rangle\nn\\
&=&
\int d^4x e^{iP\cdot x}\sum_{n,m}
\left[\sum_\lambda\int\frac{d^N p_\lambda}{(2\pi/L)^N}\right]
\langle 0|\hn_{\mu\nu}(x)\chi_n|\lambda\rangle
\langle\lambda|
h^{(m)\dagger}_{\alpha\beta}(0)\chi_m^\dagger|0\rangle~,
\end{eqnarray}
where
\begin{equation}
\chi_n\equiv M_*^N\int d^N \zbar \Omega^N \psi_n(\zbar)~.
\end{equation}
Here we have used the periodic boundary condition for $p_\lambda$ with the
spatial size $L$.
To eliminate the explicit $L$-dependence, we can rescale $\chi_n(z)$ as
\begin{equation}
\chi_n\rightarrow\frac{\chi_n}{\sqrt{L^N}}~.
\end{equation}
\par
So it is quite natural to write
\begin{eqnarray}
S&\sim&
\int d^4x\sqrt{g_4}~T^{\mu\nu} M_*^N\int d^N\zbar\sqrt{g_N}h_{\mu\nu}\nn\\
&=&
\int d^4x\sqrt{g_4}~T^{\mu\nu}\frac{1}{M_*^{1+N/2}}
\sum_n \Big[\MP\hn_{\mu\nu}(x)\Big]
\left(\frac{\chi_n}{\sqrt{L^N}}\right)~.
\end{eqnarray}
Now we can {\em define} the tensor unparticle operator $O_{\mu\nu}$ from
above as
\begin{equation}
\calO_{\mu\nu}\equiv\sum_n \MP h^{(n)}_{\mu\nu}(x)
\left[M_*^N\int d^N\zbar\Omega^N\frac{\psi_n(z)}{\sqrt{L^N}}\right]~.
\end{equation}
In this prescription, one identifies
\begin{eqnarray}
\tn_{\mu\nu}&\equiv& \MP\hn_{\mu\nu}~,\\
F_n&\equiv&M_*^N \int d^N\zbar\Omega^N \frac{\psi_n(\zbar)}{\sqrt{L^N}}~,
\label{tnFn}
\end{eqnarray}
where $F_n$ must satisfy the matching condition of Eq.\ (\ref{matchF})
(see the discussions below).
\par
\section{Discussions and conclusions}
If the KK decomposition is to be the unparticle deconstruction, one must check
whether the masses of intermediate states are vanishing, and $F_n$ defined in
Eq.\ (\ref{tnFn}) satisfies the 'decay constant' matching condition of 
Eq.\ (\ref{matchF}).
This can be easily checked by inspecting the Eq.\ (\ref{DE}).
For large $L\gg r$, $k\ll 1$ and the equation becomes
\begin{equation}
(\nabla_\zbar^2+m_\lambda^2){\hat\psi}_\lambda=0~.
\end{equation}
With the periodic boundary conditions at $\zbar=0$ and $\zbar=L$,
the solution is
${\hat\psi}_\lambda(\zbar)\sim\prod_j\sin(\pi n_{\lambda,j}\zbar_j/L)$ where
the integers $n_{\lambda,j}$ satisfy
\begin{equation}
m_\lambda^2=\sum_j \frac{\pi^2 n_{\lambda,j}^2}{L^2}~.
\end{equation}
Hence the masses are $m_\lambda\sim 1/L\to 0$ for large $L$.
\par
Also in this limit,
\begin{equation}
F_n\sim M_*^N \int d^N\zbar \frac{\psi_n(\zbar)}{\sqrt{L^N}}
\sim (p_\lambda^2)^{N/4}~,
\end{equation}
where $p_\lambda=\pi n_\lambda/L$.
Hence
\begin{equation}
\delta(P^2-p_\lambda^2)F_\lambda^2\sim (P^2)^{N/2-1}=(P^2)^{\dU-2}~,
\end{equation}
showing the proper scaling behavior.
Other coefficients can be adjusted by appropriate normalizations.
In short, the vanishing mass spectra and matching conditions guarantee our 
deconstructing methodology for unparticles.
\par
The keypoint of the equivalence between ungravity and FXD is that
the spectral density functions for both cases behave in the same way.
For ungravity, the modification of the Newtonian potential results from
\begin{equation}
V_\calU(r)\sim\int d^3{\vec p}~e^{i{\vec p}\cdot{\vec x}}\Delta_{0000}
\sim
\int d^3{\vec p}~e^{i{\vec p}\cdot{\vec x}}({\vec p}^2)^{\dU-2}
\sim\left(\frac{1}{r}\right)^{2\dU-1}~.
\end{equation}
Here the factor of $({\vec p}^2)^{\dU-2}$ originates from the unparticle
spectral density $\rho_{\mu\nu\alpha\beta}(p)$.
\par
For FXD, the modification of the potential comes from the continuum KK
excitations
\begin{equation}
V_{\rm FXD}(r)\sim\sum_{\lambda}\frac{e^{-m_\lambda r}}{r}
\sim\int dm_\lambda (m_\lambda)^{N-1}\frac{e^{-m_\lambda r}}{r}
\sim\left(\frac{1}{r}\right)^{1+N}~,
\end{equation}
which has the same power of $V_\calU$ since $1+N=2\dU-1$.
In the integration, the factor of $(m_\lambda)^{N-1}$ is the spectral density
of states in $N$ dimensions \cite{Csaki,Kingman}.
Note that
$\int dm_\lambda (m_\lambda)^{N-1}
\sim\int dm_\lambda^2(m_\lambda^2)^{N/2-1}
=\int dm_\lambda^2 (m_\lambda^2)^{\dU-2}$,
ensuring that the spectral density functions are basically the same for
ungravity and FXD.
\par
It would be quite interesting to see whether there are other ways of
realizing ungravity under different metrics.
Or, on top of it, is it always possible to realize other unparticles in the
context of FXD in general?
Although there are no explicit realizations to date for both case,
the answers are positive.
The reason is that the phase space integrations over unparticles and KK states
are very similar, as shown above.
And the deconstruction of unparticles is much like the KK mode decompositon in
extra dimensions.
Both are sum over infinite tower of states with vanishing mass gap(this is
true only for limiting case of extra dimensions),
sharing the same type of spectral density function.
For example, scalar unparticles can be easily realized within the context of
FXD developed in this work.
(One has only to ignore the spin structure and indices.)
Vector unparticles can also be incorporated with this framework, though the
spin structure might be quite different.
\par
One way of realizing unparticles in the context of deconstruction is to use
the AdS/CFT correspondence \cite{Stephanov}.
According to the AdS/CFT, for a given conformal theory in 4 dimensions
there exists a gravity theory in AdS$_5$.
In this approach, there is no connection between the scaling dimension of the
unparticle operator and the number of extra dimensions.
Rather, the 5 dimensional mass $m_5$ of the bulk scalar field is closely
related to $\dU$ via $m_5^2=\dU(\dU-4)$.
The scalar unparticle operator is defined by
\begin{equation}
\calO_\calU(x)\equiv\lim_{z\to 0}~z^{-\dU}\Phi(x,z)~,
\end{equation}
where $z$ is the fifth coordinate and $\Phi(x,z)$ is a scalar field living
in the 5 dimensions.
If both AdS$_5$ theory and FXD describe the same unparticle physics of 4
dimensions, then there must be some kind of relations between them.
But the issue is far beyond the scope of this paper.
\par
In conclusion, we have deconstructed ungravity in terms of 
extra dimensional theory, 
realizing tensor unparticles in $\AdS4N$ for the first time.
The main result is that the scaling dimension of unparticles is closely
related to the number of extra dimensions, as already claimed in the early
literatures.
In this context, unparticle physics is equivalent to the FXD theory.
Thus it is quite interesting and challenging to explore the LHC probes of
extra dimensions to see whether they can be interpreted as unparticles when
the number of extra dimensions turns out to be deviated from an integer.
And it remains also as future studies to check whether there is a deeper 
connection between unparticle and FXD (or even with AdS/CFT)
at much more fundamental level.
\begin{acknowledgments}
The author gives thanks to Jinn-Ouk Gong for helpful comments.
\end{acknowledgments}


\begin{thebibliography}{99}
\bibitem{Georgi}
 H.~Georgi,
  Phys.\ Rev.\ Lett.\  {\bf 98}, 221601 (2007)
  [arXiv:hep-ph/0703260];
Phys.\ Lett.\  B {\bf 650}, 275 (2007)
  [arXiv:0704.2457 [hep-ph]].
\bibitem{Kingman}
  K.~Cheung, W.~Y.~Keung and T.~C.~Yuan,
  Phys.\ Rev.\ Lett.\  {\bf 99}, 051803 (2007)
  [arXiv:0704.2588 [hep-ph]];
Phys.\ Rev.\  D {\bf 76}, 055003 (2007)
  [arXiv:0706.3155 [hep-ph]].
\bibitem{U}
M.~Luo and G.~Zhu,
Phys.\ Lett.\  B {\bf 659}, 341 (2008)
  [arXiv:0704.3532 [hep-ph]];
Y.~Liao,
  Phys.\ Rev.\  D {\bf 76}, 056006 (2007)
  [arXiv:0705.0837 [hep-ph]];
T.~Kikuchi and N.~Okada,
  Phys.\ Lett.\  B {\bf 661}, 360 (2008)
  [arXiv:0707.0893 [hep-ph]],
  Phys.\ Lett.\  B {\bf 665}, 186 (2008)
  [arXiv:0711.1506 [hep-ph]];
T.~Kikuchi, N.~Okada and M.~Takeuchi,
  Phys.\ Rev.\  D {\bf 77}, 094012 (2008)
  [arXiv:0801.0018 [hep-ph]];
A.~Lenz,
  Phys.\ Rev.\  D {\bf 76}, 065006 (2007)
  [arXiv:0707.1535 [hep-ph]];
N.~V.~Krasnikov,
  Int.\ J.\ Mod.\ Phys.\  A {\bf 22}, 5117 (2007)
  [arXiv:0707.1419 [hep-ph]],
  Mod.\ Phys.\ Lett.\  A {\bf 23}, 3233 (2008)
  [arXiv:0802.0830 [hep-ph]];
F.~Sannino and R.~Zwicky,
  Phys.\ Rev.\  D {\bf 79}, 015016 (2009)
  [arXiv:0810.2686 [hep-ph]];
D.~C.~Dai and D.~Stojkovic,
  Phys.\ Rev.\  D {\bf 80}, 064042 (2009)
  [arXiv:0812.3396 [gr-qc]];
J.-P.~Lee,
  arXiv:0710.2797 [hep-ph],
  arXiv:0803.0833 [hep-ph],
AIP Conf.\ Proc.\  {\bf 1078}, 626 (2009)
  [arXiv:0809.3311 [hep-ph]].
\bibitem{Goldberg}
  H.~Goldberg and P.~Nath,
  Phys.\ Rev.\ Lett.\  {\bf 100}, 031803 (2008)
  [arXiv:0706.3898 [hep-ph]].
\bibitem{Das}
  S.~Das, S.~Mohanty and K.~Rao,
  Phys.\ Rev.\  D {\bf 77}, 076001 (2008)
  [arXiv:0709.2583 [hep-ph]].
\bibitem{Mureika}
 J.~R.~Mureika,
  Phys.\ Lett.\  B {\bf 660}, 561 (2008)
  [arXiv:0712.1786 [hep-ph]];
  Phys.\ Rev.\  D {\bf 79}, 056003 (2009)
  [arXiv:0808.0523 [hep-ph]].
\bibitem{ADD}
N.~Arkani-Hamed, S.~Dimopoulos and G.~R.~Dvali,
  Phys.\ Lett.\  B {\bf 429}, 263 (1998)
  [arXiv:hep-ph/9803315];
I.~Antoniadis, N.~Arkani-Hamed, S.~Dimopoulos and G.~R.~Dvali,
  Phys.\ Lett.\  B {\bf 436}, 257 (1998)
  [arXiv:hep-ph/9804398].
\bibitem{RS}
L.~Randall and R.~Sundrum,
  Phys.\ Rev.\ Lett.\  {\bf 83}, 3370 (1999)
  [arXiv:hep-ph/9905221];
  Phys.\ Rev.\ Lett.\  {\bf 83}, 4690 (1999)
  [arXiv:hep-th/9906064].
\bibitem{Nima}
  N.~Arkani-Hamed, S.~Dimopoulos, G.~R.~Dvali and N.~Kaloper,
  Phys.\ Rev.\ Lett.\  {\bf 84}, 586 (2000)
  [arXiv:hep-th/9907209].
\bibitem{Garriga}
  J.~Garriga and T.~Tanaka,
  Phys.\ Rev.\ Lett.\  {\bf 84}, 2778 (2000)
  [arXiv:hep-th/9911055].
\bibitem{Csaki}
  C.~Csaki, J.~Erlich, T.~J.~Hollowood and Y.~Shirman,
  Nucl.\ Phys.\  B {\bf 581}, 309 (2000)
  [arXiv:hep-th/0001033].
\bibitem{Stephanov}
  M.~A.~Stephanov,
  Phys.\ Rev.\  D {\bf 76}, 035008 (2007)
  [arXiv:0705.3049 [hep-ph]].
\bibitem{Maldacena}
  J.~M.~Maldacena,
  Adv.\ Theor.\ Math.\ Phys.\  {\bf 2}, 231 (1998)
  [Int.\ J.\ Theor.\ Phys.\  {\bf 38}, 1113 (1999)]
  [arXiv:hep-th/9711200].
\bibitem{jplee}
J.~P.~Lee,
  Phys.\ Rev.\  D {\bf 79}, 076002 (2009)
  [arXiv:0901.1020 [hep-ph]].
\bibitem{GIR}
  B.~Grinstein, K.~A.~Intriligator and I.~Z.~Rothstein,
  Phys.\ Lett.\  B {\bf 662}, 367 (2008)
  [arXiv:0801.1140 [hep-ph]].
\end{thebibliography}
\end{document}